\documentclass[preprint,review,12pt]{elsarticle}

\usepackage[a4paper, left=20mm, right=20mm, top=30mm, bottom=30mm]{geometry}

\usepackage{amsmath,amssymb,amsfonts}
\usepackage{mathrsfs}

\usepackage{graphicx}
\usepackage{subfigure}
\usepackage{float}
\usepackage{tikz}

\usepackage{booktabs}
\usepackage{multirow}
\usepackage{array}
\usepackage{makecell}
\usepackage{tablefootnote}
\usepackage{threeparttable}

\usepackage{algorithm}
\usepackage{algorithmic}

\usepackage{hyperref}
\hypersetup{
    colorlinks=true,
    linkcolor=blue,
    citecolor=blue,
    urlcolor=blue
}

\usepackage{xcolor}
\usepackage{enumitem}
\usepackage{natbib}

\usepackage{siunitx}

\journal{Physica A: Statistical Mechanics and its Applications}

\begin{document}

\begin{frontmatter}

\title{Ising models on the hydrogen peroxide and other lattices} 

\author[a]{Xiaofeng Qian\corref{cor1}}
\ead{xfqian272@hotmail.com} 

\author[b]{Youjin Deng}
\ead{yjdeng@ustc.edu.cn} 

\author[c,d]{Lev~N. Shchur}
\ead{lshchur@hse.ru}

\author[a]{Henk~W.~J. Bl\"ote \fnref{fn1}} 

\fntext[fn1]{Professor Henk Bl\"ote served as the PhD supervisor of the first and second authors and made extensive contributions to this work, including finalizing the analysis and completing the manuscript draft; his intellectual guidance shaped the direction of the study and the development of its core results. Sadly, he passed away before the manuscript was brought to publication, and his contributions were substantial.} 

\cortext[cor1]{Corresponding author.}


\affiliation[a]{organization={Lorentz Institute, Leiden University},
                    addressline={P.O. Box 9506},
                    city={Leiden},
                    postcode={2300 RA},
                    country={The Netherlands}}

\affiliation[b]{organization={University of Science and Technology of China},
                    city={Hefei},
                    postcode={230027},
                    country={China}}

\affiliation[c]{organization={Laboratory for Computational Physics, HSE University},
                    city={Moscow},
                    postcode={123458},
                    country={Russia}}

\affiliation[d]{organization={Landau Institute for Theoretical Physics},
                    city={Chernogolovka},
                    postcode={142432},
                    country={Russia}}

\begin{abstract}

We perform a Monte Carlo analysis of the Ising model on many three-dimensional lattices. By
means of finite-size scaling we obtain the critical points and determine
the scaling dimensions. As expected, the critical exponents agree with the three-dimensional Ising
universality class for all models. The irrelevant field, as revealed by the correction-to-scaling amplitudes, appears
to be relatively large. Combining the Monte Carlo results for the hydrogen peroxide lattice with
those for five other three-dimensional lattices, we obtain a set of data covering a wide range of the
irrelevant temperature field. This is helpful in the determination of the parameters describing the
corrections to scaling. As a consequence, new results are obtained for the universal parameters
describing Ising criticality in three dimensions, with reduced error margins in comparison with earlier
Monte Carlo analyses. The critical exponents describing the thermodynamic singularities are
determined by the temperature renormalization exponent $y_t = 1.58693 (9)$ and the magnetic renormalization
exponent $y_h = 2.48178 (5)$. The corrections to scaling are governed by the irrelevant
exponent $y_1 = -0.821 (5)$.

\end{abstract}


\end{frontmatter}


\section{Introduction}
\label{sec:intro}

The universality class of the three-dimensional Ising model covers a wide range of models
and systems with short-range interactions and a scalar order parameter. Among these, the
critical points of simple gas-liquid systems may be considered the most common ones. It is
thus appropriate to formulate theoretical approaches to describe the phase transitions of this
type. In the absence of exact solutions, many efforts have been made to obtain the universal
constants of this class as accurately as possible. The development of the renormalization
theory~\cite{R1} did not only yield more and new insight in phase transitions, but also accurate
results for the universal exponents; see Refs.~\cite{R2,R3} and references therein. Furthermore,
the increasing number of terms in series expansions of the Ising model enables increasingly
accurate estimates of these exponents, see e.g., Ref.~\cite{R4} and references therein. In a different
type of approach, the recent availability of fast and relatively cheap computers has opened
the possibility to obtain similarly accurate estimates by means of Monte Carlo simulations
and finite-size scaling, see e.g., Ref.~\cite{R5} and references therein.
Since the latter approach will also be the subject of the present work, we outline the
ideas behind it. The finite-size scaling behavior of observables such as the specific heat,
susceptibility etc. at criticality is typically described by divergences according to power laws
modified by corrections to scaling. In first order, these corrections are proportional to the
irrelevant renormalization fields that parametrize the critical subspace of the Ising model.
The presence of such corrections naturally leads to a reduction of the accuracy of the
determination of the universal parameters. The numerical analysis of such a correction
requires the determination of its amplitude as well as of the associated irrelevant exponent.
These corrections are usually important, and neglecting them has yielded incorrect results~\cite{R6,R7,R8}.

While an accurate determination of the irrelevant exponent would favor the use of an Ising
model with a relatively large correction amplitude, i.e., a large irrelevant field, a problem may
arise in the accuracy of the numerical analysis because the quadratic term in the irrelevant
scaling field may become important, thus affecting the accuracy of the determination of the
irrelevant exponent and other universal parameters. A model with a small irrelevant field can
well be used to determine the other universal parameters, at least if the irrelevant exponent
is approximately known. However, such a model is obviously not suitable to determine the
irrelevant exponent.

The strategy followed in Ref.~\cite{R5,R9} therefore relied on the simultaneous analysis of several
models, with considerably different irrelevant fields. The availability of such Monte Carlo-generated
finite-size data enables an analysis that can be done in two stages, the first of which
is to check whether the data for the different models are consistent with universality. Then,
the second stage, which is based on the assumption that universality is \textit{exactly} satisfied,
combines the relevant Monte Carlo data for all models and applies a fitting procedure in
which the universal parameters occur only once.

The choice of the models is based on the observed dependence of the correction-to-scaling
amplitudes on further-neighbor interactions~\cite{R10} and on the introduction of a vacant spin
state as specified by the Blume-Capel model~\cite{R11,R12}. The effects of interactions with further
neighbors on the irrelevant field can be understood in terms of the crossover between the long-range
mean-field fixed point and the short-range Ising fixed point~\cite{R13}. This explains why
the variation of the number of interacting neighbors has a strong effect on the correction-to-scaling
amplitudes. Furthermore the relation between the activity of the vacancies (spin-zero
states) and the irrelevant field, which was described in the renormalization analysis of the
Potts model by Nienhuis et al.~\cite{R14}, explains the behavior of the correction amplitudes in
the Blume-Capel model.

The range of irrelevant fields between the Ising and mean-field fixed points is simply
accessible by adjusting the number of interacting neighbors. But the range of irrelevant
fields on the other side of the Ising fixed point is less easy to explore. In principle, that
range of irrelevant fields can be accessed by introducing antiferromagnetic interactions with
further neighbors. However, a sufficiently efficient algorithm for the simulation of such
models is not available, and we will focus instead on individual models that are known or
expected to lie in the desired range. One of these is the simple-cubic lattice gas~\cite{R5} with
nearest-neighbor exclusions, but it is difficult to obtain the same level of statistical accuracy
as for other models. We thus select another Ising model with few interacting neighbors,
namely the model on the hydrogen peroxide lattice, with coordination number three. On
this basis we expect that the irrelevant field will even exceed (in absolute value) that of the
Ising model on the diamond lattice~\cite{R5} which has coordination number four.

In addition to the study of the model on the hydrogen peroxide lattice, the present paper
reconsiders five of the models of Ref.~\cite{R5}  and adds new simulation results for system size
$L = 256$, as well as results for smaller systems with improved statistics. Our intention is not
only to narrow down the statistical error margins, but also to obtain data for a wide range
of values of the irrelevant field. Subsequent analysis will purportedly improve our knowledge
of the universal parameters.

The outline of the next sections is as follows. Section~\ref{sec:models} defines the six models and the
simulation methods, and supplies some further details on the simulations and the computational
resources. Part of the data for the nearest-neighbor Ising model were generated by
the Cluster Processor~\cite{R15}. We include a subsection on this special-purpose
computer. 
Section~\ref{sec:sep} provides finite-size scaling analyses for the six models separately, and the results
are subjected to a consistency test with universality. An analysis of the Ising universal
parameters by means of a simultaneous fitting procedure is performed in Sec.~\ref{sec:sim}. A short
discussion in Sec.~\ref{sec:dis} concludes this paper.

\section{Models and computational resources}
\label{sec:models}

\subsection{The model on the hydrogen peroxide lattice}

\begin{figure}
\begin{center}
\includegraphics[width=0.5\linewidth]{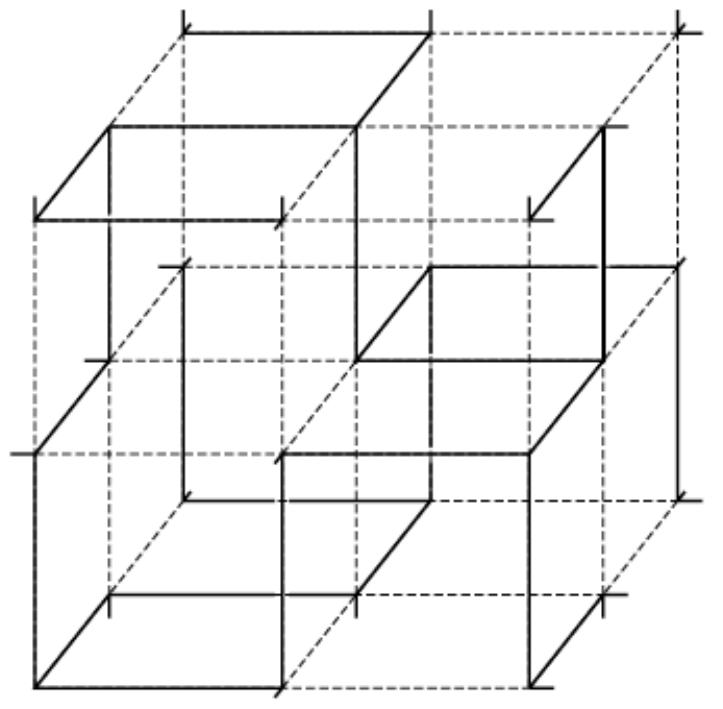}
\caption{Sketch of the hydrogen peroxide lattice. The Ising spins are located at the vertices of
the lattice. Each spin interacts with three nearest neighbors. The nearest-neighbor couplings are
shown as solid lines. The hydrogen peroxide lattice has cubic symmetry, and all sites are equivalent.
In this sketch, the lattice sites coincide with those of the simple cubic lattice, but the number of
spin-spin couplings is less by a factor of two.}
\label{fig1}
\end{center}
\end{figure}

The spin-$\frac{1}{2}$ Ising model is described by the reduced Hamiltonian
\begin{equation}
{\cal H} /k_BT = -K \sum_{<ij>} s_i s_j,
\label{eq1}
\end{equation}
where the spins $s_i$ are labeled by the lattice site number $i$. They are normalized such as to
assume values $\pm 1$, and are located on the vertices of the hydrogen peroxide lattice, which
is shown in Fig.~\ref{fig1}. The spins interact with only three of their neighbors, as shown by the
solid lines in the figure. The lattice has cubic symmetry, so that the three main crystal axes
are equivalent. The critical point of the Ising model on the hydrogen peroxide lattice has
been determined by means of series expansions by Leu et al.~\cite{R32} as $K_c = 0.573795(16)$.
This Ising model is interesting because of the low coordination number and the expected
special value of the irrelevant field, and the equivalence with a nonintersecting-loop model~\cite{R34}. 
However, as far as we were able to check, no further research on this model has been reported.

\subsection{Other models, simulations and random-number generator}

In addition to the model on the hydrogen peroxide lattice, which we label as model 1,
the present Monte Carlo analysis includes five more Ising-like models. They can be defined
in terms of the following spin-$1$ Hamiltonian
\begin{equation}
{\cal H} /k_BT = -K \sum_{i<j} \theta(R-r_{ij}) s_i s_j + D\sum_k (s_k^2-1),
\label{eq2}
\end{equation}
where the sums are over the sites of the diamond lattice (model 2), or the simple-cubic lattice
(models 3-6). The lattice sites are labeled by $i, j$ and $k$. The distance between sites $i$ and $j$ is
denoted $r_{ij}$ , and the step function $\theta(x)$, which is equal to $0$ for $x < 0$ and equal to 1 for $x\ge 0$,
defines a cutoff of the interactions of strength $K$ at range $R$. The ranges $R$ are chosen such
that interactions are restricted to the four nearest neighbors (model 2), six nearest neighbors
(models 3 and 4), 26 closest neighbors (model 5), or the 32 closest neighbors (model 6). The
spins can assume the values $s_i = \pm 1$ or 0. However, for models 1, 2, and 3 the parameter $D$ is
set to the value $D = -\infty$ so that the $s_i = 0$ state is excluded, and the model in effect reduces
to the spin-$\frac{1}{2}$ model. For model 4, we take $D = \ln 2$ which leads to a system with strongly
reduced corrections to scaling~\cite{R9}; moreover, the model can be mapped on a spin-$\frac{1}{2}$ model,
which allows efficient simulations by means of a cluster algorithm~\cite{R9}. A newer algorithm
does not use the spin-$\frac{1}{2}$ representation and applies cluster steps directly to spins 1, which
include transitions between spin-1 and 0 states~\cite{R5} without the need to include local updates
by Metropolis steps. Further details about the algorithms, also describing how models 5
and 6 can be efficiently simulated, appear in Ref.~\cite{R5}. It is noteworthy that it is possible to
apply the star-triangle transformation to model 1, which leads to another Ising model with
only half the number of spins, which may thus enable somewhat faster simulations. Since
the data sampling would then become more time-consuming, we have chosen to simulate
the original model on the hydrogen peroxide lattice.

\begin{table}[htbp]
\caption{Description of the six simulated Ising models. The abbreviation ``\# nb.'' stands for
neighbor spins coupled to each spin. $R$  indicates the maximum interaction distance between coupled spins,
expressed in the units of a cubic cell of size 2.}
\begin{center}
\begin{tabular}{l l l l c r} 
\toprule 
Model & $D$      & Lattice                      & Spin                     & \# nb.   & $R$\\
\midrule
1         & $-\infty$ & hydr. peroxide & spin-$\frac{1}{2}$  & 3  & $1$ \\ 
2         & $-\infty$ & diamond         & spin-$\frac{1}{2}$  & 4 & $\frac{\sqrt{3}}{2}$ \\ 
3         & $-\infty$ & simple cubic    & spin-$\frac{1}{2}$  & 6 &  1 \\ 
4         & $\ln 2$   & simple cubic     & spin-1                    & 6   &  1\\ 
5         & $-\infty$ & simple cubic    & spin-$\frac{1}{2}$  & 26  & ${\sqrt{3}}$\\ 
6         & $-\infty$ & simple cubic    & spin-$\frac{1}{2}$  & 32 & 2\\ 
\bottomrule
 \end{tabular}
 \end{center}
 \label{table1}
 \end{table}

The definitions of the six models are summarized in Table~\ref{table1}. For the models on the simple
cubic lattice, the unit cell contains one site and the finite-size parameter $L$ is thus equal to
the size of the periodic box, expressed in nearest-neighbor distances. The unit cells of the
hydrogen peroxide and the diamond lattices can be chosen as cubes containing eight sites.

It is thus convenient to assign a size 2 to these unit cells, so that a system of size $L$
contains $L^3$ spins, just as the models on the simple cubic lattice. Expressed in these units,
the nearest-neighbor distances are however $1$ and $\sqrt{3}/2$ respectively. Using a cubic cell
of size 2, the positions modulo 2 of the spins on the diamond lattice are (0, 0, 0), (1, 1, 0),
(1, 0, 1), (0, 1, 1), ($\frac{1}{2}, \frac{1}{2},\frac{1}{2} $), ($\frac{3}{2}, \frac{3}{2},\frac{1}{2} $), 
($\frac{3}{2}, \frac{1}{2},\frac{3}{2} $), ($\frac{1}{2}, \frac{3}{2},\frac{3}{2} $).

The Monte Carlo simulations used systems of size $L \times L \times L$ with periodic boundary
conditions, and system sizes in the range $4 \le L \le 256$. The bulk of the simulations took
place very close to the known critical points~\cite{R5}. A small fraction of the computer time was
spent on simulations somewhat further away from the critical point, in order to determine
the temperature-dependences of the sampled quantities for some smaller systems $L \le 32$.
Table~\ref{table2} presents the number of samples in units of $10^7$ taken per system size and the number
of simulation sweeps before taking each sample. The system sizes were taken in the range
$4 \le L \le 256$.

\begin{table}[htbp]
\caption{Number of samples (in units of $10^7$) and simulation steps per sample for each model. We use the
notation $M \times N$ to indicate that $10^7\; M$ samples have been taken at intervals of $N$ single-cluster
steps. Smaller even system sizes $4 \le L\le 18$ are also included in the simulations. Odd systems
$5\le L \le 15$ were simulated where possible, i.e., for models 3, 4, 5, and 6. At least $200 \times 10^7$ samples
were taken for the smaller system sizes $L < 20$.}
\begin{center}
\resizebox{\textwidth}{!}{
\begin{tabular}[width=2\columnwidth]{l c c  c c c c c c c c} 
\toprule
 Model\textbackslash L & 20 & 22 & 24 & 28 & 32 & 40 & 48 & 64 & 128 & 256   \\ 
\midrule
1 & $200{\times}5$ &  $200{\times} 6$ & $200{\times} 6$ & $200{\times}$7 & $ 200{\times} 8$ & $240{\times} 10$ & $200{\times} 12$ & $120{\times} 16$ & $ 24 {\times} 32 $ & $4.0{\times} 64$  \\ 
2 & $220{\times} 10$ &  $220{\times} 10$ & $260{\times} 10$ & $280{\times} 10$ & $ 300{\times} 10$ & $600{\times} 10$ & $160{\times} 20$ & $140{\times} 20$ & $ 20 {\times} 40 $ & $4.0{\times} 64$  \\ 
3 & $200{\times} 10$ &  $200{\times} 11$ & $320{\times} 12$ & $200{\times} 14$ & $ 240{\times} 16$ & $240{\times} 20$ & $120{\times} 24$ & $120{\times} 32$ & $ 20 {\times} 64 $ & $3.6{\times} 128$  \\
4 & $200{\times}6$ &  $200{\times} 6$ & $200{\times} 6$ & $160{\times}$6 & $ 160{\times} 8$ & $140{\times} 10$ & $100{\times} 12$ & $80{\times} 16$ & $ 24 {\times} 32 $ & $3.0{\times} 64$  \\ 
5 & $200{\times}20$ &  $120{\times} 22$ & $120{\times} 24$ & $120{\times}$28 & $ 120{\times} 32$ & $140{\times} 40$ & $100{\times} 48$ & $60{\times} 64$ & $ 16 {\times} 128 $ & $3.0{\times} 256$  \\ 
6 & $200{\times}20$ &  $120{\times} 22$ & $120{\times} 24$ & $120{\times}$28 & $ 120{\times} 32$ & $120{\times} 40$ & $100{\times} 48$ & $60{\times} 64$ & $ 16 {\times} 128 $ & $3.0{\times} 256$  \\ 
\bottomrule 
\end{tabular}
}
 \end{center}
 \label{table2}
 \end{table}

Since very long computer runs are impractical, we divided the simulations for a specified
model, system size and coupling into many independent runs. The total number of runs was
over 8000. Next, the data for a given model, system size and coupling were combined into
an average on the basis of their individual standard deviations. These standard deviations
were obtained from a partitioning of each run into two thousand subruns and subsequent
statistical analysis.

During this step of data compression, we discarded four runs that displayed unacceptable
deviations (over four standard deviations) from the average. Some of the discarded runs
could be traced back to the same computer which was thus diagnosed to be defective.

Much attention is needed in order to obtain Monte Carlo results free of biases due to
imperfect random number generators. This means actually that such biases must be reduced
to a level that is small in comparison with the statistical uncertainties. In this work we
employ pseudo-random-number generators based on binary shift registers. It is known that
such pseudo-random-numbers are not sufficiently random for some purposes. Significant
errors were reported (see, e.g., Refs.~\cite{R16, R17, R18, R19, R20, R21,R22,R23,R56,R57}). 
In many applications, the deviations from
randomness are dominated by three-bit correlations which are a direct consequence of the
production rule.

In order to avoid systematic effects we have to deal with the problem that the present
simulations are of a considerable length, and we wish to avoid tests with a similar magnitude.
For this reason we make use of the ``scalability'' of the biases as reported earlier for cluster
simulations~\cite{R23,R56}. On this basis we can extrapolate the biases found for smaller systems and
shorter shift registers to estimate the bias effects in our simulations. Additional confidence
in the validity of our approach is based on the observation~\cite{R23} that the magnitude of the
observed biases decreases rapidly as a function of the number of correlated bits. By modulo-2
addition of two maximum-length shift registers with a three-bit production rule, it is possible
to form a new sequence whose dominant correlation is a nine-bit one, which does almost
satisfy the maximum-length criterion. On the basis of the aforementioned extrapolations~\cite{R23,R56}, 
we believe that this procedure suppresses the biases below the present threshold of
observability.

\subsection{The cluster processor}

The preparations for the construction of the Cluster Processor started in 1992. The plan
was to construct one prototype processor and, after verification of its performance, to build
a system of twelve parallel units for the Wolff simulation~\cite{R24} of the three-dimensional Ising
model with linear sizes up to $L = 256$. The first stage of this project, the construction
of the prototype processor, was completed successfully in the expert hands of Dr. A. L.
Talapov, who acted as the hardware specialist. The Monte Carlo data generated by this
processor led to a few publications~\cite{R15, R25, R26}. Four of the twelve processors could be
made operational~\cite{R45}.  In spite of the delay and the less than projected capacity of the system, fairly
accurate results for the three-dimensional Ising universal quantities could be obtained~\cite{R27}.
The data generated by it contribute significantly to the numerical data for model 3 analyzed in the present work. 
For the largest system size
$L = 256$ it contributed about one third of the data for model 3, for $L = 128$ about one sixth.

\section{Separate finite-size analyses}
\label{sec:sep}

\subsection{Dimensionless ratio $Q$}

During the simulations, the magnetization density $m$ was sampled, as well as its second
moment $\langle m^2\rangle$ and its fourth moment $\langle m^4\rangle$. From these data we derived the ratio
$Q =\langle m^2\rangle^2/\langle m^4\rangle$, which is related to the Binder cumulant~\cite{R28,R29}, 
and tends to a universal constant
at criticality. The analysis of $Q$ formulated here is based on renormalization arguments.
Since both $\langle m^2\rangle$ and $\langle m^4\rangle$ can be expressed in terms of derivatives of the free energy to the
magnetic field, the scaling of the free energy yields the scaling behavior of $Q$ near the critical
point. Since the geometric factors, which contain the derivative of the magnetic scaling field
to the physical magnetic field coupled to $m$, cancel between the numerator and denominator
of $Q$, the zeroth order term in $Q$ is dimensionless. The finite-size-scaling behavior of $Q$ of
the $j$-th model follows by including the finite-size parameter $L$ in the free energy, taking the
appropriate derivatives, and renormalization to finite size 1, as
\begin{equation}
Q(t,u_j,L) = \tilde Q(L^{y_t}t,\, L^{y_1}u_j) +\sum_k b_{k,j} L^{y_k}.
\label{eq3}
\end{equation}

The temperature field which is denoted $t$, and the irrelevant field denoted $u_j$ , are dependent
on the nonuniversal parameters appearing in the Hamiltonian Eq.~\eqref{eq2}, while the function $\tilde Q$
and the exponents $y_t$ and $y_1$ are universal. A few correction terms on the right are
included, of which one is due to the analytic background term in the second derivative of
the free energy to the magnetic field, whose leading term is of order $L^{2y_h- d}$, where $d{=}3$ is
the dimensionality and $y_h$ is the magnetic renormalization exponent. This effect leads to a
correction exponent $y_2{=} d {-} 2y_h$. Furthermore the temperature scaling field may depend
in second order on the physical magnetic field, so that higher order derivatives of the free
energy to the magnetic field will include corrections with finite-size exponent $y_3 {=} y_t {-}2y_h$.
In order to bring Eq.~\eqref{eq3} in a suitable form to describe Monte-Carlo generated finite-size
data for $Q$, we expand the parameters $t$ and $u$ in the physical fields, namely the couplings
$K_j$:
\begin{eqnarray}
\label{eq4}
t&=&a_j(K_j-K_{cj})+c_j (K_j-K_{cj})^2 + \cdots \\ \nonumber 
u&=&u+ v_j(K_j-K_{cj})+ \cdots,  
\end{eqnarray}
where $K_{cj}$ is the critical point of the $j$-th model. The coefficients are nonuniversal and also
carry a label indicating the model. The universal function $\tilde Q$ is expanded as
\begin{eqnarray}
\label{eq5}
\tilde Q(L^{y_t}t,\,L^{y_1}u) &=& Q^{(0,0)} + L^{y_t}  t Q^{(1,0)} + L^{2y_t}  t^2 Q^{(2,0)} \\ \nonumber 
&+& L^{3y_t}  t^3 Q^{(3,0)} + L^{4y_t}  t^4 Q^{(4,0)} + \cdots \\ \nonumber
&+& L^{y_1}  u Q^{(0,1)} + L^{2y_1}u^2 Q^{(0,2)} + \cdots \\ \nonumber
&+& L^{y_t+y_1}  tu Q^{(1,1)} + \cdots , 
\end{eqnarray}
where the superscripts of the expansion coefficients $Q^{(k,l)}$ denote derivatives of $\tilde Q$ to its
arguments. They are thus universal. Combination of Eq.~\eqref{eq3} and the two expansions yields, Eqs.~\eqref{eq4} and \eqref{eq5},
a result of which the leading terms are
\begin{eqnarray} 
\label{eq6}
Q&=&Q^{(0)}+q_{1j} (K_j-K_{cj})L^{y_t} +q_{2j} (K_j-K_{cj})^2L^{2y_t} \\ \nonumber
&+&q_{3j} (K_j-K_{cj})^3L^{3y_t}+q_{4j} (K_j-K_{cj})^4L^{4y_t}  \\ \nonumber
&+& c_j (K_j-K_{cj})^2L^{y_t} + b_{1j}L^{y_1} + b_{2j}L^{y_2} + b_{3j} L^{y_3}   \\ \nonumber
&+&  d_j(K_j-K_{cj}) L^{y_t+y_1},
\end{eqnarray}
where we abbreviated $Q^{(0,0)}$ as $Q^{(0)}$. The constants $q_{ij}$ are the product of universal and
nonuniversal constants from both expansions. These as well as other free parameters in
Eq.~\eqref{eq6} were estimated by means of least-squares fits to the Monte Carlo data, for each
of the six models. Our first aim is now to determine the universal quantity $Q^{(0)}$. To this
purpose we set $y_1 {=} {-} 0.82(3), y_2 {=} {-}1.963(3)$, and $y_3 {=}{-}3.375(3)$ as already specified in
earlier analyses of $Q$ in Refs.~\cite{R9,R27}, and $y_t$ was taken to be $1.587$~\cite{R4,R5}.

\begin{table}[htbp]
\caption{Separate fits of the dimensionless ratio $Q$ for the six models. Columns 2 to 4 show
the estimated critical coupling, the asymptotic value of $Q$ and the amplitude of the finite-size
correction due to the irrelevant field, with the irrelevant exponent fixed at $y_1{=}-0.82$. Some
data for small system sizes, namely $L<6,6,6, 8, 10$ and 10 for models 1 to 6 respectively, were
excluded from these fits. Columns 5 and 6 show the value of the irrelevant exponent when left
free, and the corresponding value of $Q$. This did not lead to meaningful results for model 4. The
minimum system sizes used for these fits were $L{=} 4$ for models 1, 2 and 3, $L{=} 5$ for model 5, and
$L{=} 8$ for model 6. The error estimates shown between parentheses are one standard deviation.}
\begin{center}

\begin{tabular}[width=2\columnwidth]{l l l S S l}
\toprule
 Model& $K_c$  & $Q^{(0)}$ &  {$b_1$} & {$y_1$} & $Q_{free}$  \\
\midrule
1 & 0.57371376(6) & 0.62339(4)  & 0.1183(7) & -0.83(2) & 0.6235(1) \\
2 & 0.36973981(4) & 0.62341(3)  & 0.1139(6) & -0.85(2) & 0.6236(1) \\
3 & 0.22165458(2) & 0.62346(2)  & 0.0930(4) & -0.82(2) & 0.6235(1) \\
4 & 0.39342218(4) & 0.62349(4)  & -0.0033(5) & {-}  &  {-} \\ 
5 & 0.043038238(4) & 0.62336(3) & -0.1112(5) & -0.76(2) & 0.6239(2) \\
6 & 0.034326876(6) & 0.62327(6) & -0.216 (1) & -0.75(2) & 0.6240(3) \\ 
\bottomrule
 \end{tabular}
 \end{center}
 \label{table3}
 \end{table}

The data for some small system sizes were excluded from these fits because additional
corrections are troublesome in such small systems, as becomes apparent from the residual 
$\chi^2$. The results for $K_{cj}$ , $Q^{(0)}$ and $b_{1j}$ as obtained from these fits are shown in Table~\ref{table3}.
In addition we show results of fits with $y_1$ as a free parameter, for $y_1$ as well as for the
resulting value of the universal ratio denoted $Q_{free}$. The statistical errors are shown between
parentheses as one standard deviation in the last decimal place.

The agreement among the results in Table~\ref{table3} for the universal quantities $Q$ and $y_1$ is
obviously less than perfect. The results for $Q^{(0)}$ and $y_1$ display a systematic dependence
on the amplitude $b_1$, which is a measure of the irrelevant field. For $Q^{(0)}$ this dependence is
obviously nonlinear. The detection of this effect is a result of the higher statistical accuracy
of our data in comparison with previous work~\cite{R5}. The type of dependence is suggestive
of the second-order term in the expansion of $Q(t, u, L)$ to $u$, i.e., we have to add a term
$\frac{1}{2}L^{2y_1}uQ^{(0,2)}$ to Eq.~\eqref{eq5} which amounts to including a term $b_{4j}L^{2y_1}$ in Eq.~\eqref{eq6}. Results of
fits including this term, and with $y_1 {=}{-}0.82$ fixed, are shown in Table~\ref{table4}. They are better
behaved, as visible from the  $\chi^2$ criterion, and from the absence of a systematic dependence
of the results for $Q^{(0)}$ as a function of $b_1$.

The results in the third column in Table~\ref{table4} confirm that the six systems fit well in one common Ising
universality class. Due to the additional parameter, the uncertainty margins are larger than
those in Table~\ref{table3}, but still smaller than those of the results of Ref.~\cite{R5}.

\begin{table}[htbp]
\caption{Separate fits of the dimensionless ratio $Q$ for the six models, with a quadratic term in
the irrelevant field included. Smallest system sizes included in these fits are $L = 4, 4, 4, 4, 6$, and
8 for models 1 to 6 respectively. The last column shows the amplitude of the leading correction
to scaling. It reflects the magnitude of the irrelevant field. The error estimates shown between
parentheses are one standard deviation.}
\begin{center}
\begin{tabular}[width=2\columnwidth]{l l l S}
\toprule
 Model& $K_c$  & $Q^{(0)}$ & {$b_1$}    \\
\midrule
1 & 0.57371378(7) & 0.62343(8) & 0.117(3) \\
2 & 0.36973981(4) & 0.62350(7) & 0.110(3) \\ 
3 & 0.22165458(2) & 0.62348(5) & 0.093(2) \\
4 & 0.39342218(5) & 0.62343(9) & -0.000(3) \\
5 & 0.043038244(4) & 0.62352(6) & -0.117(2) \\
6 & 0.034326877(5) & 0.62357(15) & -0.232(7) \\
\bottomrule
 \end{tabular}
 \end{center}
 \label{table4}
 \end{table}

\begin{figure}[htbp]
\center
\includegraphics[width=0.5\linewidth]{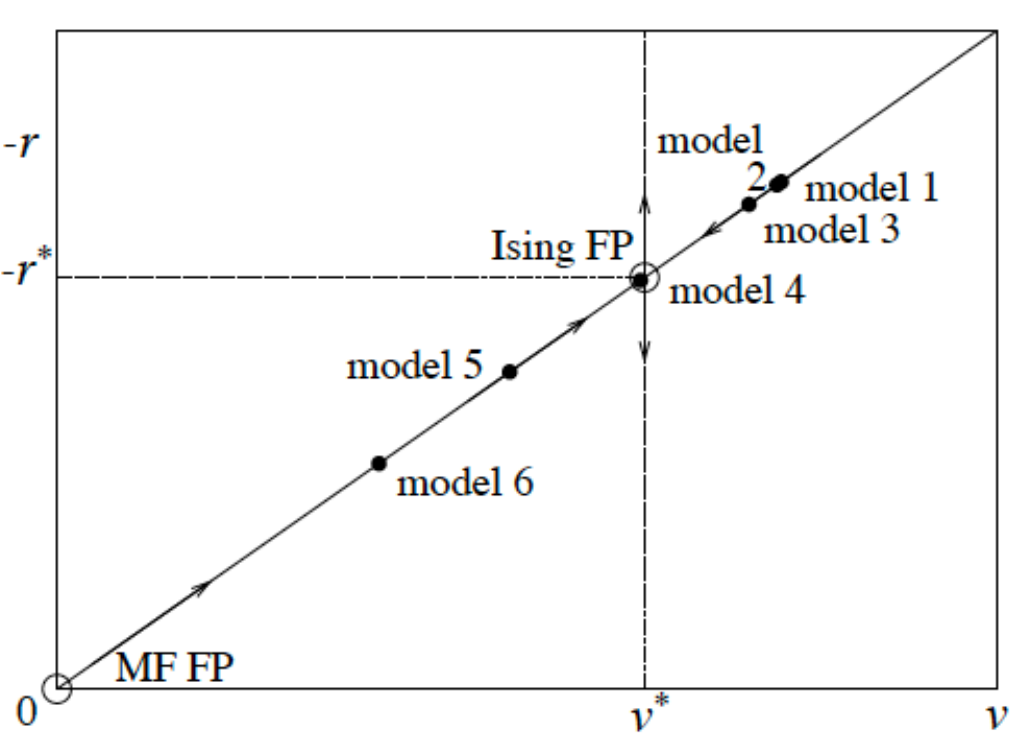}
\caption{Sketch of positions of models 1-6 in the parameter space $(r, v)$ of the $\phi^4$ model, where $r$ is
a temperature-like parameter, and the $\phi^4$ amplitude $v$ parametrizes the irrelevant field of the Ising
model. The positions of the six models are shown by black circles ( \scalebox{1.2}{$\bullet$} ). The mean-field and Ising fixed points, 
which are shown as open circles ( \scalebox{2.5}{$\circ$} ), are located at $(-r, v){=}(0, 0)$ and $(-r^*, v^*)$, respectively.}
\label{fig2}
\end{figure}

Also included in Table~\ref{table3} and ~\ref{table4} are the amplitudes of the corrections due to the irrelevant field
of these models. Up to a constant factor these results represent the actual values of the
irrelevant scaling field in first order. As mentioned in Ref.~\cite{R5}, these values reflect the positions of the
critical points of these systems under a renormalization mapping on the Landau-Ginzburg-Wilson 
Hamiltonian of the $\phi^4$ model, which is
\begin{equation}
{\cal H}(\phi)/k_BT=\int d {\bf x} \left[r\phi^2({\bf x} )  +v\phi^4 ({\bf x} ) +\nabla^2 \phi ({\bf x} )  \right],
\label{eq7}
\end{equation}
where $\bf x$ denotes the space coordinates. The square-gradient term represents the Ising
particle-particle interaction, while the parameter $r$ is temperature-like, and $v$ parametrizes
the irrelevant field. Renormalization analysis of such a system in less than four spatial
dimensions~\cite{R1} shows two fixed points, one of which is located at $(r,v){=}(0, 0)$ and represents the
mean-field fixed point, while the other sits at nonzero values $(r^*<0, v^* > 0)$ and represents
the Ising fixed point. The crossover scaling function for the Binder ratio~\cite{R30} provides a
scale for the interval between both fixed points, as parametrized by the Ising irrelevant field.
The determination of the amplitudes of the irrelevant corrections in $Q$ thus determine the
positions of the six models in the $(r^*<0, v^* > 0)$ diagram shown in Fig.~\ref{fig2}.

\subsection{Separate analyses of the susceptibility}

Differentiation of the finite-size scaling equation for the free energy density shows that
the magnetic susceptibility $\chi$  behaves as
\begin{equation}
\chi(t,v,L) = x(t) + L^{2y_h-d} \left( \frac{\partial h}{\partial  H}\right)^2 \chi(L^{y_t}t,L^{y_1}v,1),
\label{eq8}
\end{equation}
where the function $\chi$ on the right hand side is universal in its two remaining arguments.
The contribution $x(t)$ is determined by the second derivative of the analytical part of the
free energy density with respect to the physical magnetic field $H$. The dependence of
the magnetic scaling field $h$ on $H$ is not universal and may, in first order, be denoted as
$h=\sqrt{w_j}H$. Taylor-expansion of Eq.~\eqref{eq8} yields

\begin{eqnarray}
\label{eq9}
\chi&=& x_j + s_j(K_j-K_{cj}) \\ \nonumber 
&+& L^{2y_h-d} \left[ p_{j0} +p_{j1} (K_j-K_{cj})L^{y_t} \right.\\ \nonumber
&+&  p_{j2} (K_j-K_{cj})^2L^{2y_t}  + p_{j3} (K_j-K_{cj})^3L^{3y_t}\\ \nonumber
&+& p_{j4} (K_j-K_{cj})^4L^{4y_t}+\sum_k b_{kj} L^{y_t} \\ \nonumber
&+&\left. d_{j} (K_j-K_{cj}) L^{y_t+y_1} \right],
\end{eqnarray}
where we have used the expansion of the temperature field as in Eq.~\eqref{eq4}. The constants $p_{jk}$
denote the product of universal expansion parameters of the function $\chi$ and nonuniversal
constants including the $w_j$. The results of the fits for the six models are given in Table~\ref{table5}.

\begin{table}[htbp]
\caption{Separate fits of the susceptibility $\chi$ with the irrelevant exponent fixed at $y_1{=}-0.82$. 
The minimum system sizes used for this fit are $L = 4, 4, 6, 4, 5, $ and 8 for models 1 to 6 respectively. 
Columns 2–5 list the estimated critical coupling, the coefficient of the leading term in the temperature-field expansion, 
the magnetic renormalization exponent $y_h$, and the amplitude $b_1$ of the finite-size correction associated with the irrelevant field. 
The error estimates shown between parentheses are one standard deviation.}
\begin{center}
\begin{tabular}[width=2\columnwidth]{l l l l S} \toprule
 Model& $K_c$  & $p_0$ & $y_h$ & {$b_1$}   \\ \midrule
1 & 0.57371380(7) & 2.026(3) & 2.48161(14) & -0.547(13) \\
2 & 0.36973982(5) & 1.756(3) & 2.48156(19) & -0.460(17) \\
3 & 0.22165459(2) & 1.556(2) & 2.48160(15) & -0.351(9) \\
4 & 0.39342224(5) & 0.934(1) & 2.48168(16) & 0.004(6) \\
5 & 0.043038242(5) & 0.990(1) & 2.48167(13) & 0.257(6) \\
6 & 0.034326874(6) & 0.878(10) & 2.48137(28) & 0.428(13) \\ \bottomrule
 \end{tabular}
 \end{center}
 \label{table5}
 \end{table}

One of the variations of the fit formula included quadratic terms in the irrelevant field.
This led to slightly larger values of $y_h$ for five of the six models, although not significant in
individual cases.

\subsection{Separate analysis of $Q_p$}

During the Monte Carlo simulations, also the nearest-neighbor sum $e$
\begin{equation}
e=\left<S_{nn}\right> = \sum_{(nn)} \langle s_is_j\rangle 
\label{eq10}
\end{equation}
was sampled. This quantity represents the interaction energy density for models 1, 2 and 4.
For models 5 and 6, the nearest-neighbor sum is not the interaction energy, but it is still a
quantity that scales as the energy.

Combination with the magnetization data allows the Monte Carlo sampling of the derivative
$Q_p \equiv \partial Q/\partial K_{nn}$ of the Binder ratio $Q$ with respect to the nearest-neighbor interaction.
This derivative is different from that to $K$ for models 5 and 6. It is determined by the
sampled quantities as
\begin{eqnarray}
\label{eq11}
Q_p&=&2\frac{\langle m^2S_{nn}\rangle}{\langle m^2\rangle}-\frac{\langle m^4 S_{nn}\rangle}{\langle m^4\rangle} - \langle S_{nn}\rangle \\ \nonumber 
&=& \frac{1}{Q} \frac{\partial Q}{\partial t} \frac{\partial t}{\partial K_{nn}}.
\end{eqnarray}
The quantity $Q_p$ is known~\cite{R5, R9, R27} to yield fairly accurate results for the thermal exponent
$y_t$. According to Eq.~\eqref{eq3}, the quantity $Q_p$ behaves near the critical point as
\begin{equation}
Q_p(t,v,L) =  L^{y_t}  \frac{\partial t}{\partial  K_{nn}}  Q(L^{y_t}t,L^{y_1}v,1),
\label{eq12}
\end{equation}
Expansion of $t$ and $\partial t/\partial  K_{nn}$, and of the function $Q_p$ in its two remaining arguments, yields
\begin{eqnarray}
\label{eq13}
Q_p&=& L^{y_t}  \left[  r_{j0} +r_{j1} (K_j-K_{cj})L^{y_t}\right.\\ \nonumber 
&+&  r_{j2} (K_j-K_{cj})^2L^{2y_t}  + r_{j3} (K_j-K_{cj})^3L^{3y_t}\\ \nonumber
&+& r_{j4} (K_j-K_{cj})^4L^{4y_t} + b_{1j}L_{y_1} + d_j L^{y_2} \\ \nonumber
&+&\left. c_{j} (K_j-K_{cj})  \right],
\end{eqnarray}
where we denote contributions due to the analytic part of the free energy as $d_j L^{y_2}$ . The
leading contribution is seen to diverge with $L$ as $L^{y_t}$. This is a much stronger divergence
than that of the specific heat $C$ which displays only weakly divergent behavior as $L^{2y_t-d}$.
Therefore, $Q_p$ is more suitable than $C$ to determine the temperature exponent $y_t$. The
free parameters in Eq. ~\eqref{eq13} were fitted according to the least-squares criterion. The critical
points were fixed at the values listed in Table~\ref{table5}. The results are shown in Table~\ref{table6}. They
include the quantity $b_1/r_0$ which is proportional to the irrelevant field and may thus be
compared with Tables~\ref{table4} and~\ref{table5}.

\begin{table}[htbp]
\caption{Separate fits of the derivative $Q_p$ of the Binder ratio with the irrelevant exponent fixed
at $y_1{=}-0.82$. The minimum system sizes used for this fit are $L = 6, 6, 7, 6, 7,$ and 6 for models
1 to 6 respectively. The errors quoted between parentheses are one standard deviation.}
\begin{center}
\begin{tabular}[width=2\columnwidth]{l l l S } 
\toprule
 Model& $r_0$  & $y_t$ & {$b_1/r_0$}  \\ 
\midrule
1 & 0.5302(7) & 1.58674(30) & -0.082(6) \\
2 & 0.8227(8) & 1.58689(21) & -0.074(4) \\
3 & 1.3529(14) & 1.58710(22) & -0.057(5) \\
4 & 1.0627(13) & 1.58678(28) & -0.008(5) \\
5 & 1.4224(16) & 1.58713(24) & 0.043(5) \\
6 & 1.4365(13) & 1.58720(20) & 0.086(3) \\
\bottomrule
 \end{tabular}
 \end{center}
 \label{table6}
 \end{table}

\section{Simultaneous finite-size scaling analyses}
\label{sec:sim}

\subsection{Simultaneous analysis of $Q$}

The results of Sec.~\ref{sec:sep} indicate that the six models under investigation belong to the
same universality class. We thus assume that the universal parameters of these six models
are exactly equal and we analyze the combined numerical data for $Q$ of these systems
simultaneously. Separation of the universal and nonuniversal constants in Eqs.~\eqref{eq4} and~\eqref{eq5}
yields
\begin{eqnarray} 
\label{eq14}
Q&=& Q^{(0)}\\ \nonumber
&+&Q^{(1)} a_j (K_j-K_{cj})L^{y_t} +Q^{(2)} a^2_j(K_j-K_{cj})^2L^{2y_t} \\ \nonumber
&+& Q^{(3)} a^3_j(K_j-K_{cj})^3L^{3y_t}+Q^{(4)} a^4_j(K_j-K_{cj})^4L^{4y_t}  \\ \nonumber
&+& c_j (K_j-K_{cj})^2L^{y_t} + b_{1j}L^{y_1} + b_{2j}L^{y_2} + b_{3j} L^{y_3}   \\ \nonumber
&+&  d_j(K_j-K_{cj}) L^{y_t+y_1},
\end{eqnarray}
where we abbreviated $Q^{(k,0)}$ as $Q^{(k)}$ and combined the other $Q^{(k,l)}$ with their nonuniversal
prefactors into constants denoted $c_j$ and $b_{1j}$. The term with amplitude $c_j$ is due to the
nonlinear dependence of the temperature field $t$ as a function of $K_j$. The correction exponents
$y_i$ have the same meaning as above. On the basis of Eq.~\eqref{eq14}, least squares fits were
applied to the combined data for $Q$ of the six models. The exponents were set at the values
given above, except the irrelevant one $y_1=y_i$ which is now treated as a free parameter.
Furthermore we set $Q^{(1)} = 1$, which defines a scale of the temperature field.

In this type of simultaneous fit, the universal constants $Q^{(k)}$ and the exponents appear
only once, so that the total number of free parameters is reduced appreciably. Also the
effect mentioned in Sec.~\ref{sec:intro}  contributes to the accuracy of the fit, in particular to that of the
irrelevant exponent $y_1$. The results, which are summarized in Table~\ref{table7}, do indeed display a
substantially improved accuracy in comparison with those of the separate fits.

\begin{table}[htbp]
\caption{Simultaneous fit of the ratio $Q$. The minimum system sizes that were included in
this fit were $L{=} 6$ for models 1-5 and $L {=}8$ for model 6. The error margins quoted between
parentheses are two standard deviations.}
\begin{center} 
\begin{tabular}[width=\columnwidth]{c c c c c c} \toprule
$y_1$ &  $Q^{(0)}$ &  $Q^{(1)}$ & $Q^{(2)}$ & $Q^{(3)}$ & $Q^{(4)}$ \\ 
-0.821(5) &  0.62356(5) & 1 (fixed) &  0.85(3) & $-3.2(2)$& $ -6.0 (27)$ \\ \midrule
$K_c^{(1)}$ &  $K_c^{(2)}$ &  $K_c^{(3)}$ & $K_c^{(4)}$ & $K_c^{(5)}$ & $K_c^{(6)}$ \\ 
0.57371386(9) & 0.36973988(6) & 0.22165461(3) & 0.39342221(6) & 0.043038246(7) & 0.034326877(5) \\ \midrule
$a_1$ & $a_2$ & $a_3$ & $a_4$ & $a_5$ & $a_6$ \\
0.329(2) & 0.5115(14) & 0.844(4) & 0.664(2) & 4.04(4) & 4.98(5) \\ \midrule
$b_{11}$ & $b_{12}$ & $b_{13}$ & $b_{14}$ & $b_{15}$ & $b_{16}$ \\
0.1124(17) & 0.1088(17) & 0.0897(14) & $-0.0050(9)$ & $-0.119(4)$ & $-0.233(8)$ \\ \bottomrule
 \end{tabular}
 \end{center}
 \label{table7}
 \end{table}

\subsection{Simultaneous analysis of $\chi$}

We rewrite the parameters $p_{jk}$ in the right-hand side of Eq.~\eqref{eq9} as the product of universal
expansion parameters $\chi^{(k)}$ of the function $\chi$ in its first argument, and nonuniversal expansion
parameters $a_j$ . Then Eq.~\eqref{eq9} changes into
\begin{eqnarray} 
\label{eq15}
\chi&=& x_j + s_j (K_j-K_{cj}) +L^{2y_h-d}w_j \left[ \chi^{(0)} \right. \\ \nonumber 
&+& \chi^{(1)} a_j (K_j-K_{cj}) L^{y_t} + \chi^{(2)} a^2_j (K_j-K_{cj})^2 L^{2y_t} \\ \nonumber
&+& \chi^{(3)} a^3_j (K_j-K_{cj})^3 L^{3y_t}+\chi^{(4)} a^4_j (K_j-K_{cj})^4 L^{4y_t} \\ \nonumber
&+& \left. b_jL^{y_j} + c_j (K_j-K_{cj}) L^{y_t+y_1} + \cdots \right],
\end{eqnarray}
where we have expanded the temperature field as $t = a_j(K_j - K_{cj}) + b_j(K_j - K_{cj})^2 + \cdots $.

Fits were made to the Monte Carlo data for $\chi$ on the basis of Eq.~\eqref{eq15}. The results are
presented in Table~\ref{table8}. Just as for $Q$, the scale of the temperature and magnetic scaling
fields may be chosen freely so that we may fix two parameters, namely $\chi^{(0)}{=} \chi^{(1)}  {=} 1$ in
Eq.~\eqref{eq15} during the fit. Again the reduction of the number of free parameters is accompanied
by an improved accuracy of the results in comparison with the separate fits.

\begin{table}[htbp]
\caption{Simultaneous fit of the magnetic susceptibility $\chi$ with the temperature exponent fixed
at $y_t{=}1.587$ and the irrelevant exponent at $y_1{=}-0.82$. Also the first two expansion parameters
were fixed at $\chi^{(0)}{=}1$ and $\chi^{(1)}{=}1$. Errors are quoted as one sigma.}
\begin{center}
\begin{tabular}[width=\columnwidth]{c c c  c c c}
\toprule
$y_h$ &  $\chi^{(0)}$ &  $\chi^{(1)}$ & $\chi^{(2)}$ & $\chi^{(3)}$ & $\chi^{(4)}$ \\ 
2.48178(5) & 1 (fixed) & 1 (fixed) & $-0.13(2)$& 0.82 (fixed) & 1.587 (fixed) \\
\midrule
$K_c^{(1)}$ &  $K_c^{(2)}$ &  $K_c^{(3)}$ & $K_c^{(4)}$ & $K_c^{(5)}$ & $K_c^{(6)}$ \\ 
0.57371386(4) & 0.36973989(3) & 0.22165462(2) & 0.39342222(3) & 0.043038244(3) & 0.034326881(3) \\
\midrule
$w_1$ & $w_2$ & $w_3$ & $w_4$ & $w_5$ & $w_6$ \\
2.023(1) & 1.7530(8) & 1.5509(7) & 0.9328(4) & 0.9886(5) & 0.8746(4) \\
\midrule
$a_1$ & $a_2$ & $a_3$ & $a_4$ & $a_5$ & $a_6$ \\
1.31(1) & 1.989(3) & 3.28(1) & 2.606(5) & 16.0(2) & 19.4(1) \\
\bottomrule
 \end{tabular}
 \end{center}
 \label{table8}
 \end{table}

\subsection{Simultaneous analysis of $Q_p$}

Separation of the universal parameters following from the expansion of the universal
function $Q_p$ in Eq.~\eqref{eq12} and the nonuniversal ones yields
\begin{eqnarray} 
\label{eq16}
Q_p&=& L^{y_t} p_j \left[ Q_p^{(0)} + Q_p^{(1)} a_j (K_j-K_{cj})L^{y_t} \right.\\ \nonumber
&+& Q_p^{(2)} a^2_j (K_j-K_{cj})^2L^{2y_t} + Q_p^{(3)} a^3_j (K_j-K_{cj})^3L^{3y_t}  \\ \nonumber
&+& Q_p^{(4)} a^4_j (K_j-K_{cj})^4L^{4y_t} + b_j L^{y_1}  + d_j  L^{y_t-y_1}  \\ \nonumber
&+& \left. e_j (K_j-K_{cj})L^{y_1+y_t} +  c_j (K_j-K_{cj}) \right],
\end{eqnarray}
where the universal parameters $Q_p^{(k)}$ are determined by the function $Q_p$. The nonuniversal
parameters $p_j$ are defined as $\partial t / \partial K_{nn}$ for the $j$-th model. As above we took the critical points
from Table~\ref{table7}  as fixed parameters in the fitting procedure. The results of these least-squares
fits of Eq.~\eqref{eq16} to the Monte Carlo results appeared to depend only weakly on the precise
value of the critical points. The results are shown in Table~\ref{table9}.

We performed several variations of this procedure. For instance we included a term
proportional to $L^{y_1}(K_j-K_{cj})$ in the square brackets of Eq.~\eqref{eq16}. However, this did not lead
to a clear reduction of the residual $\chi^2$ of the fit. The uncertainty margins shown in Table~\ref{table9}
include the variations due to different procedures, and the dependence on the cutoff at small
system sizes.

\begin{table}[htbp]
\caption{Result of simultaneous fit of $Q_p$. One sigma errors are shown between parentheses.}
\begin{center} 
\begin{tabular}[width=\columnwidth]{c c c c c c}
\toprule
$y_t$ &  $Q_p^{(0)}$ &  $Q_p^{(1)}$ & $Q_p^{(2)}$ & $Q_p^{(3)}$ & $Q_p^{(4)}$ \\ 
1.58693(9) & 1 (fixed) & 0.1 (fixed) & $-9(2)$ & $-7(3)$ & 53(25) \\
\midrule
$a_1$ & $a_2$ & $a_3$ & $a_4$ & $a_5$ & $a_6$ \\
$-0.35(4)$ & $-0.54(7)$ & $-0.89(11)$ & $-0.71(9)$ & $-4.4(5)$ & $-5.2(6)$\\ 
\midrule
$p_1$ & $p_2$ & $p_3$ & $p_4$ & $p_5$ & $p_6$ \\
0.5298(2) & 0.8227(3) & 1.3543(6) & 1.0622(5) & 1.4238(6) & 1.4391(6) \\
\bottomrule
 \end{tabular}
 \end{center}
 \label{table9}
 \end{table}

\section{Discussion}
\label{sec:dis}

By means of an analysis of extensive Monte Carlo simulations of six three-dimensional
Ising-like models, we have succeeded in refining the results for several universal constants
of the Ising universality class. Furthermore, the margins for deviations from universality
between the six models are reduced, which is significant because a rigorous basis for universality
is still lacking. The present confirmation of universality is in line with earlier findings~\cite{R5,R31}.

Furthermore the results obtained in the simultaneous fit allow an additional confirmation
of universality. As noted above, the derivative $a_j$ of the temperature field to the coupling of
the $j$-th model occurs in the expansion of $Q$ as well as in that of $\chi$. They differ only because
they are expressed in different units. Thus, according to universality, the ratios between the
amplitudes $a_j$ of $Q$, $\chi$ and $Q_p$ should be model-independent. Table~\ref{table10} accordingly displays the
$a_j^{(Q)}/a_j^{(\chi)}$ and $a_j^{(Q)}/a_j^{(Q_p)}$ for the six models. The results are in a good agreement with
universality. Table~\ref{table10}  includes the ratio $a_j^{(Q)}/p_j^{(Q_p)}$, which may expected to be constant only
for models 1 to 4. This is because the amplitudes $p_j$ are defined through differentiation with respect to the nearest-neighbor coupling $K_{nn}$, 
whereas for models 5 and 6 the coupling $K$ differs from $K_{nn}$ due to the presence of both nearest‑ and next‑nearest‑neighbor interactions.

\setlength{\tabcolsep}{1pt}
\begin{table}[htbp]
\caption{Results for the amplitude ratios $a_j^{(Q)}/a_j^{(\chi)}$ , $a_j^{(Q)}/a_j^{(Q_p)}$ and $a_j^{(Q)}/p_j^{(Q_p)}$, where the superscripts
refer to the expansion parameters used in the fits for $Q$, $\chi$ and $Q_p$. Theory predicts entries
to be equal on each line, except for models 5 and 6 on line 3.}
\begin{center}
\begin{tabular}[width=\columnwidth]{l S S S S S S}
\toprule
Model &  {1} & {2} & {3} & {4} & {5} & {6}  \\
\midrule
$a_j^{(Q)}/a_j^{(\chi)}$ &  0.253(3) & 0.257(1) & 0.257(1) & 0.255(1) & 0.253(3) & 0.257(3) \\
$a_j^{(Q)}/a_j^{(Q_p)}$ &  -0.93(10) & -0.94(11) & -0.95(11) & -0.94(11) & -0.92(11) & -0.96(12) \\
$a_j^{(Q)}/p_j^{(Q_p)}$ &  0.621(4) & 0.622(2) & 0.623(3) & 0.625(2) & 2.84(3) & 3.46(4) \\ 
\bottomrule
 \end{tabular}
 \end{center}
 \label{table10}
 \end{table}

The improvement of the results in comparison with Ref.~\cite{R5} is based on better statistical
accuracies, and simulation of larger systems of size $L = 256$. The six models were selected
such that the irrelevant field covers a wide range. On the one side of this range we have
the model on the hydrogen peroxide lattice with only 3 nearest neighbors, and on the other
side of the scale we have the equivalent-neighbor model on the simple-cubic lattice with
four layers of neighbors, which is, as illustrated in Fig.~\ref{fig2}, relatively close to the mean-field
model in terms of crossover between the mean-field and Ising fixed points. In agreement
with earlier findings, the simultaneous analysis of the Monte Carlo data helps to reduce the
uncertainty margins, especially for the errors of the irrelevant exponent $y_1$ and of quantities,
like the Binder ratio $Q$, that are sensitive to the value of $y_1$.

The reduction of the error margins in our final results is not a simple and direct result of
the better statistics of the Monte Carlo data. Reduction of the statistical errors can reveal
new corrections to the leading scaling behavior that were below the detection threshold in
earlier analyses. For instance, this applies to the correction term with exponent $y_1 - y_t$ in
the analysis of $Q_p$. Since the addition of new free parameters in the fit formula typically
increases the statistical error margins of the results, the improvement of the results is less
than expected on the basis of statistics only. Another problem with a similar effect is that
the cutoff at small systems may have to be chosen at a larger value of $L$.

Our results are summarized in Tables~\ref{table11} and~\ref{table12}. These tables also 
include some existing results for easy comparison. 

More recent estimates of the critical point for model 3, based on Monte Carlo renormalization‑group and finite‑size‑scaling analyses~\cite{R52, R50, R53}, are listed in Tables~\ref{table11}, where they are seen to be consistent with our results.

Remarkably, the recent advances in the conformal bootstrap have yielded the most precise determinations to date 
of the leading scaling dimensions in the three‑dimensional Ising universality class. 
In particular, the works of Kos \emph{et al.}~\cite{newR1}, Simmons‑Duffin~\cite{newR2}, 
and Chang \emph{et al.}~\cite{newR3} 
provide exceptionally accurate estimates for the scaling dimensions of the temperature, 
magnetic‑field, and leading irrelevant operators, denoted by 
$\Delta_{\epsilon}$, $\Delta_{\sigma}$  and $\Delta_{\epsilon'}$, 
respectively. Using the standard relations
$y_t = 3 - \Delta_{\epsilon}$,  
$y_h = 3 - \Delta_{\sigma}$, 
and 
$y_1 = 3 - \Delta_{\epsilon'}$,
we convert these scaling dimensions into the corresponding critical exponents. The resulting 
values are now included in Table~\ref{table12} to facilitate direct comparison with our results. 
The small discrepancy observed in comparison with our results may point to methodological 
constraints intrinsic to Monte Carlo simulations or to a possible underestimation of 
the error bars reported in our analysis.

\setlength{\tabcolsep}{3pt}
\begin{table}[htbp]
\caption{Comparison between present results and existing values for the critical points. Errors quoted are two sigma.}
\begin{center}
\begin{threeparttable}
\resizebox{\columnwidth}{!}{
\begin{tabular}{l l l l l l l l}
\toprule
Model & 1 & 2 & 3  & 4 & 5 & 6 \\
\midrule
$K_c$ (present)           & 0.57371385(9)   & 0.36973987(6)     & 0.22165459(3)  & 0.39342220(6)  & 0.043038246(7) & 0.034326877(5) \\
$K_c$ (Ref.~\cite{R5}) & -                   & 0.36973980(9)     & 0.22165455(5)  & 0.39342225(9)  & 0.0430381(4)   & 0.03432687(2) \\
$K_c$ (other refs.)   & 0.573795(16)~\cite{R32} & 0.36978(4)~\cite{R35}  & 0.221655(2)~\cite{R37}    & 0.3934220(7)~\cite{R9}  & 0.0430381(4)~\cite{R30}  & 0.03432685(15)~\cite{R30} \\
$K_c$ (other refs.)  & -  & 0.3697(8)~\cite{R36} & 0.22165463(8)~\cite{R49}   & 0.39342239(8)~\cite{R49}  & 0.0432~\cite{R38} & - \\
$K_c$ (other refs.)  & 0.5737138(4)\tnote{a} ~\cite{R46} & - &  0.221654631(8)~\cite{R53}   & -  &- & - \\
$K_c$ (other refs.)  & -  & - &     0.221654626(5)~\cite{R50}  & -  &- & - \\
$K_c$ (other refs.)  & -  & - &   0.2216547(1)~\cite{R52}   & -  &- & - \\
\bottomrule
 \end{tabular}}

\footnotesize
\begin{tablenotes}
  \item[a]{Derived via $K_c = artanh(x_c)$ using $x_c = 0.5180815(3)$~\cite{R46}.}
\end{tablenotes}
\end{threeparttable}
 \end{center}
 \label{table11}
 \end{table}

While most of our new results agree well with those of Ref.~\cite{R5}, the result for $Q^{(0)}$ differs
more than the combined error bounds. This is mostly due to the second-order term in the
expansion of $Q$ in the irrelevant field which is now included. This term also explains the
difference of the new result $y_h = 2.48178 (5)$ with that of Ref.~\cite{R5}, which slightly exceeds the
combined error margins. It is quite satisfactory to observe that the present results for the
Ising universal parameters, obtained by different methods as renormalization, series expansions
and Monte Carlo, coincide within narrow margins. Also the results for nonuniversal
quantities, such as the critical points as obtained by different methods are, in most cases,
in an excellent agreement. There is a small discrepancy of our result $K_c = 0.57371385 (9)$
for the critical point of the hydrogen peroxide lattice model with the series-expansion result~\cite{R32}
$K_c = 0.573795 (16)$, which may be related to the large value of the irrelevant field in this
model. Our result is in accurate agreement with a very recent independent result obtained
by Liu et al.~\cite{R46} using a worm algorithm. More serious discrepancies exist with results of
a conjectured exact solution of the three-dimensional Ising model by Zhang~\cite{R44}. His result
for the critical point of the simple-cubic Ising models lies several orders of magnitude outside
the mutually consistent estimated error margins of independent results~\cite{R5}. In this
context it relevant to mention that Wu et al.~\cite{R47}, Perk~\cite{R48, newR4}, and Fisher et al.~\cite{newR5}
pointed out several fundamental errors that invalidate all its main result claimed in this approach. 

\begin{table}[htbp]
\caption{Comparison between the present results for the renormalization exponents and the
universal ratio $Q^{(0)}$ for the three-dimensional Ising universality class and literature values. The
methods to obtain these results are abbreviated as follows: RG -- renormalization of $\phi^4$ model;
HTE -- high-temperature series expansion; MC -- Monte Carlo and finite-size scaling; MCRG --
Monte Carlo renormalization; CAM -- coherent-anomaly method; CB -- conformal bootstrap.}
\begin{center}
\begin{threeparttable}
\resizebox{\columnwidth}{!}{
\begin{tabular} {l l l  l S l l}
\toprule
Refs. &  {Year}                       & {$y_t$} & {$y_h$} & {$y_1$} &  {$Q^{(0)}$} & Method \\
\midrule
Le Guillou et al. \cite{R2}         & 1980 & 1.587(4) & 2.485(2) & -0.79(3) & &  RG \\  
Nickel and Rehr \cite{R39}       & 1990 & 1.587(4) & 2.4821(4) & -0.83(5) & & HTE \\ 
Nickel \cite{R40}                     & 1991 & 1.587 & 2.4823  & -0.84 & & HTE \\ 
Baillie et al. \cite{R41}             & 1992 & 1.602(5) & 2.4870(15) & -0.825(25) & & MCRG \\ 
Landau \cite{R42}                   & 1994 & 1.590(2) & 2.482(7) & &  & MC \\ 
Kolesik et al. \cite{R43}           & 1995 & 1.586(4) & 2.482(4) & & & CAM \\ 
Bl\"ote et al. \cite{R9}             & 1995 & 1.587(2) & 2.4815(15) & -0.82(6) & 0.6233(4) & MC \\ 
Bl\"ote et al.  \cite{R10}          & 1996 & 1.585(2) & 2.4810(10) & & & MCRG \\
Guida et al. \cite{R3}              & 1998 & 1.586(3) & 2.483(2) & -0.799(11) &  & RG  \\ 
Bl\"ote et al. \cite{R27}           & 1999 & 1.5865(14) & 2.4814(5)  & -0.82(3) & 0.62358(15) & MC \\ 
Campostrini et al. \cite{R4}     & 2002 &1.5869(4) &2.48180(15) & -0.82(5) & & HTE \\ 
Deng and Bl\"ote \cite{R5}      & 2003 & 1.5868(3) & 2.4816(1) & -0.821(5) & 0.62341(4) & MC  \\ 
Butera and Comi \cite{R37} & 2005 & 1.5870(5) & 2.4818(4) & & & HTE \\ 
Hasenbusch \cite{R49}           & 2010 & 1.58725(25)\tnote{f} & 2.481865(50)\tnote{a} & -0.832(6) & & MC \\ 
El-Showk et al. \cite{R54}      & 2012 & 1.587(1)   & 2.4818(3)&  & & CB \\ 
Kos et al. \cite{newR1}          & 2016 & 1.587375(10)\tnote{b} & 2.4818511(10)\tnote{c} & & &  CB\\
Simmons‑Duffin \cite{newR2} & 2017 & 1.587375(10)\tnote{b}  &  2.4818511(10)\tnote{c}   &-0.82968(23)\tnote{d}  &  &  CB\\
Ferrenberg et al. \cite{R50}     & 2018 & 1.58752(22) & 2.48195(6) &  &  & MC \\ 
Hasenbusch \cite{R51}           & 2021 & 1.58739(7) & 2.481858(20)\tnote{a} & -0.825(20) & 0.62360(2)\tnote{e}& MC \\ 
Ron et al. \cite{R52}              & 2021 & & 2.4824(1) & & & MCRG \\ 
Chang et al. \cite{newR3}       & 2025 & 1.58737472(29)\tnote{b} & 2.481851194(24)\tnote{c} &  -0.82951(61)\tnote{d} &  &  CB\\
Present\tnote{g}                    & 2011 & 1.58693(9)  & 2.48178(5) & -0.821(5) & 0.62355(5) & MC \\ 
\bottomrule
\end{tabular}}

\footnotesize
\begin{tablenotes}
\item[a]{Derived via $y_h = (d + 2 - \eta)/2$ using $d =3 $ and $\eta =0.03627(10)$~\cite{R49}, $\eta =0.036284(40)$~\cite{R51}}.
\item[b]{Derived via $y_t=3-\Delta_{\epsilon}$ using $\Delta_{\epsilon} =1.412625(10)$ \cite{newR1,newR2}, $ \Delta_{\epsilon}=1.41262528(29)$ \cite{newR3}.}
\item[c]{Derived via $y_h=3-\Delta_{\sigma}$ using $\Delta_{\sigma} =0.5181489(10)$ \cite{newR1,newR2}, $ \Delta_{\sigma}=0.518148806(24)$ \cite{newR3}.}
\item[d]{Derived via $y_1=3-\Delta_{\epsilon'}$ using $\Delta_{\epsilon'} =3.82968(23)$ \cite{newR2}, $\Delta_{\epsilon'} =3.82951(61)$ \cite{newR3}.}
\item[e]{Estimated from $Q = 1/{U_4}$ using the critical Binder cumulant $U_4^{*} =1.60359(4)$ \cite{R51}.}
\item[f] {Derived via $y_t = 1/{\nu}$ using $\nu =0.63002(10)$~\cite{R49}.}
\item[g] {The analyses and the initial manuscript for this work were essentially completed in 2011.}
\end{tablenotes}
\end{threeparttable}
 \end{center}
 \label{table12}
 \end{table}

\section{Acknowledgments}

We are indebted to Dr. A. L. Talapov for his contributions to the Cluster Processor in
the early stages of this project, and to Dr. J. R. Heringa for valuable discussions. The
construction of the Cluster Processor was funded by the NWO (``Nederlandse Organisatie
voor Wetenschappelijk Onderzoek'') via grants  047-13-210 and  047-003-019. YD was supported by National Natural Science Foundation of China (Grant No. 12275263) and the Natural Science Foundation of Fujian Province of China (Grant No. 2023J02032).
LNS was supported by the Russian Science Foundation (Grant 25-11-00158).

\end{document}